\documentclass[10pt,twocolumn]{article} 
\usepackage{simpleConference}
\usepackage{times}
\usepackage{graphicx}
\usepackage{gensymb}
\usepackage{amssymb}
\usepackage{amsmath}
\usepackage{url,hyperref}
\usepackage{subfigure}
\usepackage[framed=true]{matlab-prettifier}

\begin{document}

\title{Open-source objective-oriented framework for head-related transfer function modeling}

\author{Adam Szwajcowski \\
\\
AGH University of Science and Technology \\ Department of Robotics and Mechatronics
\\
szwajcowski@agh.edu.pl  \\
}

\maketitle
\thispagestyle{empty}

\begin{abstract}
Throughout last 30 years, numerous head-related transfer function (HRTF) models have been developed and there are more to come. This paper describes a framework based on objective-oriented programming paradigm, in which each HRTF representation method can be implemented as a separate class. Its modular structure allows the source code to be conveniently shared between researchers, while common interface provides easy access to data regardless of the internal structure of the classes. The paper discusses difficulties of designing the framework, maintaining the balance between its flexibility and finding common features of every possible directivity representation. Exemplary use cases are included and explained. Adoption of the framework will enhance possibilities of accuracy comparison between various HRTF models, thus improving the evaluation of current and future representation methods. The framework, developed in the form of a \textsc{MATLAB} toolbox, is designed to handle not only HRTFs but also other types of spatial data, such as e.g. sound source directivity, microphone directivity, etc.
\end{abstract}

\section{Introduction}
\label{s:introduction}

With ongoing development of augmented and virtual reality technology, the need for efficient binaural rendering is increasing year by year. To convincingly reproduce spatial sound as a binaural signal, one needs to apply appropriate head-related transfer functions (HRTFs), which can vary substantially for different individuals. HRTFs, as any directivity functions, are multidimensional, covering dependency of natural acoustic filters of torso, head and pinna on direction, frequency and distance. As such, they are relatively complex datasets and there have been numerous attempts on designing efficient HRTF representations. Initially, the research focused mainly on investigating different approaches to modeling frequency part as filters of either finite or infinite impulse responses \cite{Kulkarni1995}\cite{Blommer1997}\cite{Huopaniemi1999}. In 1998, Evans et al. utilized spherical harmonics (SHs) to model direction dependency, which later on became widely acknowledged as the default method of obtaining continuity in space not only for HRTFs but also for other types of directivity \cite{Evans1998}\cite{Ziegler2017}\cite{Szwajcowski2021}. With SHs being established as the most popular method of representing space dependency, some models introduced continuous frequency and distance representations on top of SHs to create fully continuous functional models \cite{Zhang2009a}\cite{Zhang2010}\cite{Zhang2015}. Despite wide acceptance of SH approximation, there are still attempts to dethrone them, proposing alternative spherical functions, such as Slepian functions or spherical wavelets \cite{Bates2015}\cite{Hu2019}\cite{Liu2019}. Independently, there has been research going on focusing purely on the compression of HRTFs and ignoring the continuity in any dimension \cite{Kistler1992}\cite{Huang2009}\cite{Shekarchi2013}\cite{Arevalo2020}.

Due to the multidimensionality of directivity data, the results are very inconvenient to be presented comprehensively within research papers and only averaged reproduction error values are usually provided instead. What is more, the averaging might be applied using a wide variety of formulae and the evaluation can be performed on different datasets. All those variables lead to the issue where, looking at the results of two papers describing different models, one cannot tell which of them provides better accuracy due to different evaluation methods. More often than not, the only way to reliably compare HRTF models, is to reproduce them based on the mathematical description and apply evaluation method of authors' choice \cite{Brinkmann2018}\cite{Engel2022}.

Alternatively, instead of reproducing the methods, their original developers could share the source code or data files. However, there is no standardized structure for such files, which would make it tedious to learn how to link each of them with custom datasets or evaluation methods; SOFA (Spatially Oriented Format for Acoustics), the most common format for storing such type of data, includes only a few well-known representations, which are hard-coded and new models can only be added by releasing a new version of the format \cite{Majdak2013}\cite{SOFA2020}. On the other hand, for a new representation to be acknowledged and put into such a format, it has to be properly tested before. An alternative approach is to employ an open format with a modular structure, allowing for an unlimited number of representations. Some experiments with a framework utilizing such format have already been performed; however, practical application of the preliminary version of the framework indicated a need for improvements in some areas \cite{Szwajcowski2021a}. The changes were implemented and the most prominent ones are discussed later in this work.

The framework presented in this paper aims to, on one hand, improve the experience of researchers developing new HRTF models, and on the other, make the ready representations easy to share with others in order to enhance the comparability of the models. Section \ref{s:assumptions} lays down assumptions and requirements for the framework. Section \ref{s:implementation} describes the details of how these features were incorporated in the resulting toolbox. In section \ref{s:example}, the model evaluation possibilities are highlighted based on an exemplary HRTF representation. Section \ref{s:discussion} discusses advantages and limitations of proposed approach and section \ref{s:conclusion} summarizes the work.

\section{Assumptions}
\label{s:assumptions}

\subsection{Motivation}
\label{s:motivation}

%
To circumstantiate the requirements set for the framework, it is important to provide some background on common issues arising during research concerning development of new HRTF representation methods. The main problem is the difficulty in comparison to the state of the art due to wide variety of evaluation methods adapted by various researchers. Alongside model-specific advantages (e.g. continuity in space for SH-based methods), the common evaluation point is the reproduction accuracy, i.e. how much different is the directivity function after encoding and decoding applying a given model when compared to the original data? The results depend both on the choice of the evaluation data and on adopted error measure.


In the case of HRTFs, one of the most popular datasets for evaluation are the data from the original measurements of artificial head KEMAR performed at MIT (e.g. \cite{Zhang2010}\cite{Huang2009}\cite{Shekarchi2013}) \cite{Gardner1995}. Some authors also use KEMAR data but acquired from different measurements or simulated based on its 3D model (e.g. \cite{Arevalo2020}\cite{Brinkmann2018}\cite{Chen1995}) . However, even datasets derived from the same artificial head can differ substantially due to measurement imperfections and differences in its setup \cite{Andreopoulou2013}. Another popular approach is to evaluate the models on HRTFs from human subjects, either obtained from open databases (e.g. \cite{Zhang2015}\cite{Zhang2009}\cite{Kulkarni2004}) or measured specifically for the purpose of the given research (e.g. \cite{Kulkarni1995}\cite{Evans1998}\cite{Iida2007}); it is worth noting that the last approach also allows for subsequent listening tests performed by the original subjects. Finally, analytical derivations of HRTFs based on a spherical head model were also employed in some research \cite{Bates2015}\cite{Zhang2009}.

Furthermore, there is no single established error measure for HRTF reproduction accuracy. More often than not, some sort of an objective evaluation is employed that is supposed to predict the subjective perception. The error is usually averaged over the sphere and plotted against frequency or averaged over both space and frequency and given as a single value. However, there are numerous approaches to calculating the errors for individual datapoints, including different scales, norms and normalizations. The most common error measure in recent papers, however, is simple root-mean-square (RMS) of differences in dBs for consecutive directions and/or frequencies (e.g. \cite{Liu2019}\cite{Arevalo2020}\cite{Li2021}). An alternative to these relatively simple measures are performance predictions based on perceptual models, which are often quite complex (e.g. \cite{Huopaniemi1999}\cite{Brinkmann2018}\cite{Engel2022}. Finally, objective evaluation can be replaced or supplemented by a subjective one carried out utilizing different listening test procedures (e.g. \cite{Kulkarni2004}\cite{Iida2007}\cite{Li2021}). For other types of directivity, other modifications to error measure might also apply such as area-weighting for equiangular grids for sound source directivity approximation \cite{Szwajcowski2021}.

This diversity in model evaluation methods causes problems with comparability of the reproduction accuracy. For the results to be comparable, the models should be evaluated on the same data and using the same error measure. Since there is little overlapping in both of these elements, the numerical results are close to useless.
Researchers' opinions on what is the best evaluation procedure still vary significantly, as it is difficult to provide sufficient proofs of superiority of one method over the others. For this reason, instead of trying to standardize an objective evaluation method, the goal of this work was to focus on providing a framework for easier sharing of various HRTF representation, so that any model could be easily evaluated using a dataset and an error measure of user's choice. Ultimately, by enhancing the ease of handling different HRTF representations, this framework can also contribute to finding the optimal evaluation procedure. However, even if such procedure was standardized, the usefulness of the framework will not fade away, as there will always be a need for a convenient environment for development and reproduction of HRTF models, e.g. to arrange listening tests or to investigate a specific phenomenon which cannot be observed in averaged reproduction errors.

\subsection{Core features}
\label{s:core}

Based on above-described issues and difficulties, four core features required from the framework were extracted and explained below. In order to effectively handle HRTF models at the development stage, all of them need to be satisfied.

%
%
The first core feature is \textbf{flexibility} in the sense that the framework should be capable of handling just any HRTF model (or even any directivity model) possible regardless of its definition. There are models based on logarithmic and linear values, continuous and discrete in different dimensions, frequency- and time-based, etc. Despite some of the modeling approaches being preferred, the framework should allow for any imaginable representation, as long as it is meant to store directivity data.

On the other hand, each HRTF representation still serves a common purpose and so it should be possible to read the data in the same way regardless of underlying source code. Thus, a \textbf{common interface} is necessary for convenient handling of all the models.

Another important aspect is \textbf{modularity}. As described in Introduction, there have been many different models developed and more are still coming, so it is impossible to implement all of them in a single release. The most plausible solution is to create a modular directivity format, so that each user can contribute to it with their own representation method. The source code could be then shared between researchers and they should only need the core of the framework to be able to operate on the shared files.

Finally, \textbf{fixed conventions and units} are recommended to make the interface clear. On one hand, possibility of saving data using different units or coordinate systems might be plausible for the authors of the data, but on the other, it can cause inconveniences during accessing and interpreting the data. It is thus advisable to operate using a strictly defined unit and coordinate system convention, while making sure that there are available tools for conforming to this convention. Underlying functions could still operate in different coordinate systems or units, but they need to be properly connected to the interface using the globally acknowledged convention.

\subsection{Available HRTF formats}
\label{s:formats}

Throughout the years, multiple different formats of HRTF data storage were developed, from custom MAT formats designed by various database creators, to binary ones such as SDIF, OpenDAFF, and, finally, SOFA \cite{Majdak2013}\cite{Hrauda2013}\cite{Diemo2000}\cite{Wefers2010}. However, the goal for all these formats was to conveniently handle measurement data and, while successful at their intended purpose, they can only support a very limited number of hard-coded representations. Thus, such formats could not be used at the development stage as they fail in flexibility and modularity.

Another solution is the implementation of HRTFs in ITA toolbox \cite{Dietrich2012}. There, one can find a class \texttt{itaHRTF} for handling HRTFs and so each HRTF dataset can be saved as an \texttt{itaHRTF} object. While building new representation classes based on \texttt{itaHRTF} is not impossible, such approach would go against the goal of making the whole process of developing and sharing new HRTF representations easy and convenient, as the toolbox is designed around a relatively strict structure of \texttt{itaHRTF}. It is thus clear that a new way to handle HRTFs was needed to be designed from scratch in order to satisfy all features listed in section \ref{s:core} and remain simple enough to be attractive for its potential users.

\section{Implementation}
\label{s:implementation}

After carefully reflecting upon the required features, an objective-oriented approach was chosen for the task. The framework was written in the form of a \textsc{Matlab} toolbox under the name \textsc{ooDir} standing for Objective-Oriented Directivity, to highlight its universality in handling all directivity data, not only HRTFs. The toolbox alongside necessary documentation can be found at \url{https://oodir.sourceforge.io}. Below, the general design choices are described and circumstantiated. For detailed syntax description of all classes and functions, the reader is advised to read the documentation available in the repository or directly in the source files.

\begin{figure}[h]
\centering
\includegraphics[width=0.9\linewidth]{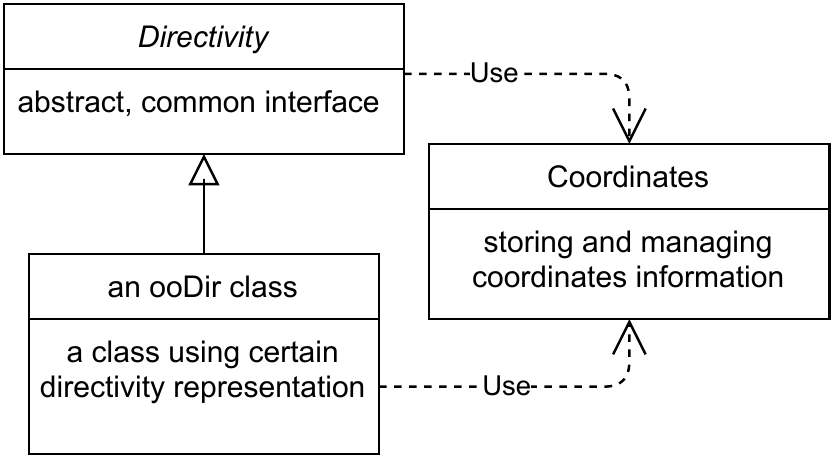}
\caption{Class scheme of the \textsc{ooDir} framework}
\label{f:scheme}
\end{figure}

The core part of the framework are two classes: an abstract superclass \texttt{Directivity} serving as the interface, and \texttt{Coordinates}, a class for handling the data coordinates. On the base of these two, subsequent classes can be built, each responsible for a specific directivity representation method. The scheme of the framework is depicted in Fig. \ref{f:scheme}. The core of the toolbox is rounded out by \texttt{RawIRs} class designed to store SOFA-like data, i.e. impulse responses for discrete coordinates in space. Since it is the most common type of directivity representation, it can be considered basic and is commonly used for deriving more advanced models and then as an accuracy benchmark for them. However, \texttt{RawIRs} is not essential for the framework to function properly and can be replaced by other classes designed for storing raw discrete data e.g. as wideband frequency data instead of impulse responses. All above mentioned classes are described in details in the following subsections.

\subsection{Directivity class}
\label{s:directivity}

The heart of the framework is the \texttt{Directivity} class, which is an abstract class, i.e. it forces a certain structure of its subclasses but cannot be instantiated itself. \texttt{Directivity} also defines some functions which are uniform for each directivity representation.

There are two properties that every \textsc{ooDir} class needs to include. One of them is \texttt{info}, a public string for general information about a given object. Since for different representation method different information might be desirable, no classical metadata fields are provided. The object creators can input all the important notes in the \texttt{info} string or provide necessary description in some other form. The motivation for such solution is the fact that during the development stage, filling all the metadata fields might be skipped, while at the stage of file sharing, the object creators can be trusted to determine what kind of information is important to attach to the datafile and in what form. Most of what is usually put in metadata in other formats (such as e.g. SOFA) is covered within properties defined specifically for a given \texttt{Directivity} subclass and \texttt{info} is only intended for information which is irrelevant to proper handling the \textsc{ooDir} object, such as e.g. ID number of HRTF set.

The other property common for each \textsc{ooDir} class is \texttt{coords}, an object of class \texttt{Coordinates} with protected access (more details in next subsection). All the remaining properties of individual \textsc{ooDir} classes are specific for employed directivity representations, but \texttt{info} and \texttt{coords} can be defined for any model and thus are required\footnote{\texttt{info} can be left as an empty string, although providing even some very basic description is recommended.}.

\texttt{Directivity} has one abstract method which is \texttt{getDataM} for reading directivity data in the form of a matrix. The way of decoding data can be different for different \textsc{ooDir} classes, but it will always have the same inputs: coordinates from which the data is to be read and requested type of data. In the early version of the toolbox, only logarithmic magnitude values were used, but, to increase the flexibility of the framework, the available types of data are now defined within each \textsc{ooDir} class (see sections \ref{s:rawirs} and \ref{s:example} for examples) \cite{Szwajcowski2021a}. The output of \texttt{getDataM} is also always the same - it is a 3D matrix of values for different directions, frequencies (or time samples when reading impulse responses) and distances in consecutive dimensions. Optionally, \texttt{getDataM} can also return true coordinates at which the data were read; these may differ from the requested ones because of coercion (for more details see section \ref{s:coordinates}).

\texttt{Directivity} also includes non-abstract methods which are uniform for each subclass. The simplest of them is \texttt{getCoords}, which simply returns the \texttt{Coordinates} object, a protected property of each Directivity subclass. This way, this property can be easily read, but cannot be modified from outside of the class to prevent corruption. Another simple method is \texttt{getDataV}, which reads the data in the form of a matrix using \texttt{getDataM} and flattens it to a vector. The other non-abstract methods are plotting methods, one for plotting spectra and one for plotting directivity balloons. The plotting functions create figures and invoke private methods depending on the continuity in frequency and in direction, for spectra and balloons respectively. Therefore, object of any Directivity subclass can invoke \texttt{plotSpectrum} and \texttt{plotBalloon} methods which subsequently call private methods depending on the object continuity in the respective dimension. These private methods, thanks to the uniform interface of \texttt{getDataM}, work the same for all \textsc{ooDir} classes. Optionally, handles to created figures can be returned to allow customization of the plot appearance.
\subsection{Coordinates class}
\label{s:coordinates}

\begin{figure}[h]
\centering
\includegraphics[width=0.65\linewidth]{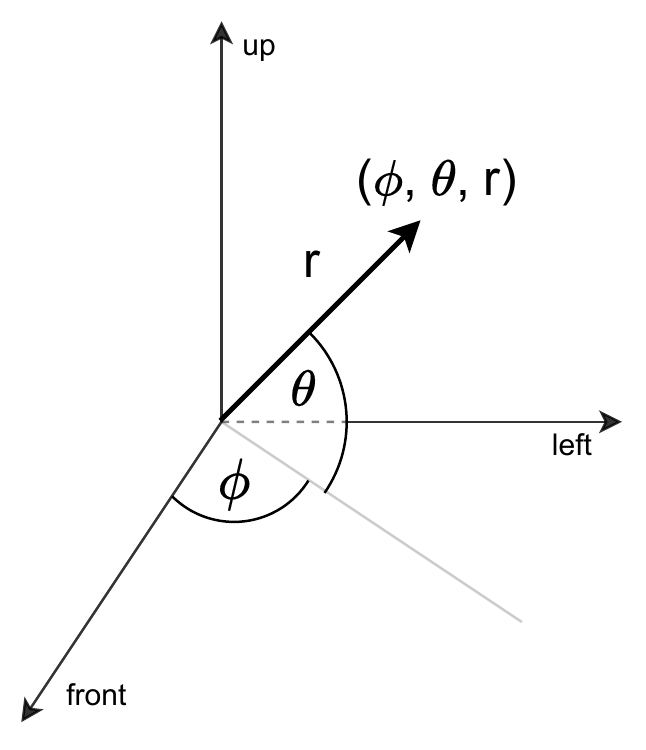}
\caption{Coordinate system used in the \textsc{ooDir} framework: $\phi$ denotes azimuth and $\theta$ denotes elevation}
\label{f:coordsystem}
\end{figure}

Even though directivity models can have abstract mathematical definitions, at the end of the day, they still need to be somehow linked to the physical coordinates. \texttt{Coordinates} is a class for holistic handling of the information on coordinates. Its properties include direction matrix as well as frequency and distance vectors; the direction matrix is compounded of two vectors, one for azimuth and elevation each, effectively being a vector of direction duplets. Azimuth is the horizontal angle, starting from 0 at the front, going anti-clockwise through left, back and right side of the center to reach 360$\degree$ at the front again, while elevation ranges from -90$\degree$ at the bottom to +90$\degree$ at the zenith (Fig. \ref{f:coordsystem}). Frequency is given in Hertz and distance in meters;  however, it is worth noting that, since near-field models are quite rare, the toolbox was designed in such a way that the distance dependency can be ignored altogether. In such cases, its value defaults to 1 m, which is already considered far-field. 

Another property of \texttt{Coordinates} is \texttt{continuity}, a three-element boolean vector for information whether the object represents continuous or discrete direction, frequency or distance, respectively. For continuous coordinates, the property vectors store only the lower and upper limits of the variable instead of its discrete values. For example, most HRTF sets lack data for low elevations and so the limits for direction-continuous representations can be imposed in such a way, that only data above certain elevation is available (even though technically it might be possible to read the data beyond this limit).

As far as the methods are concerned, \texttt{Coordinates} includes a constructor (simple reading object properties from arguments) and two public methods. One of them is \texttt{getM} which formats the vectors and outputs them as matrices of direction, frequency and distance values which are of the same size and compatible with \texttt{getDataM} called using the given \texttt{Coordinates} object as an argument. The other public method is \texttt{coerce}, which, when called on an object A with an object B as an argument, returns an object C, which is the same as object B, but with values in its vectors coerced to the nearest values in respective vectors in object A. The \texttt{coerce} makes reading the data more convenient and can be seen as a sort of nearest-neighbour interpolation. For example, when requesting a plot of a directivity balloon at 1000 Hz, one would probably be satisfied with data for 1013 Hz as well if there was no data available precisely at 1000 Hz. It can also help avoid issues arising from numerical errors. For continuous dimensions, the data is coerced only if it lies outside of the specified limits.

\subsection{RawIRs}
\label{s:rawirs}

Almost all directivity models are derived from raw measurement data in the form of discrete impulse responses (as stored in SOFA). Since this representation is so common, \texttt{RawIRs} class was provided as a basic part of the toolbox. However, directivity models can also be sometimes based on measurements in wide frequency bands (e.g. \cite{Szwajcowski2021}), in which case a similar class would have to be developed to handle such input data.

Besides the inherited properties, \texttt{RawIRs} includes also a matrix of impulse responses as well as their sampling rate in Hertz. The class constructor passes the arguments to the properties and calculates available frequency values from the length of impulse responses and their sampling rate. In the early version of the framework, the basic raw measurement representation class operated only on logarithmic-magnitude frequency values, but here, data in the form of impulse responses is used \cite{Szwajcowski2021a}. Even though nowadays magnitude spectra are often used as a starting point for directivity models (e.g. \cite{Hu2019}\cite{Li2021}), maintaining original impulse responses increases flexibility, while the frequency data can still be easily obtained by means of Fourier transform. Alternatively, the impulse responses could be replaced by their complex spectra.

\texttt{RawIRs} has no class-specific methods apart from its constructor, which simply creates an object and assigns the input data to its properties. However, it does include its own implementation of \texttt{GetDataM}, as it is only defined as abstract in the superclass. The implementation of \texttt{getDataM} for \texttt{RawIRs} allows for a wide variety of output data type, including original impulse responses, complex spectra and magnitude spectra of different forms (linear magnitude, power spectrum and logarithmic magnitude). The method coerces the coordinates from which the data are requested and then reads proper values either directly (for impulse responses) or after performing certain processing (for frequency spectra).

\subsection{Utilities}
\label{s:utilities}

Even though the three above-described classes are sufficient to make the toolbox fully functional, some additional functionalities could further improve the experience of using it. After performing some research utilizing the preliminary version of the \textsc{ooDir} framework \cite{Szwajcowski2021a}, I noticed that some routines are repeatable and thus providing tools for automating them would be desirable.

\subsubsection{SOFA converter}
\label{s:SOFA2RawIRs}

One of such routines is creating new \texttt{RawIRs} objects from measurement data. Since currently most of HRTF datasets can be found in SOFA format, I developed a converter \texttt{SOFA2RawIRs}, which loads up a SOFA file and returns a \texttt{RawIRs} object\footnote{If a SOFA file contains more than one directivity functions, a vector of \texttt{RawIRs} objects is returned instead.} based on provided data. The converter is designed to work with SOFA files using default coordinate systems and units and thus it might require adjustments to read some files using non-standard definitions.

\subsubsection{Exemplary object for testing}
\label{s:kemar}

In order to facilitate testing of new \textsc{ooDir} classes, an example dataset is needed. As listed in section \ref{s:motivation}, the most common dataset for HRTF model evaluation are the original KEMAR measurement data and so a \texttt{RawIRs} object based on these data is provided in the toolbox saved in the file \texttt{KEMAR.mat}. The object  contains data for left ear for the large pinna setup. The dataset is provided for test purposes and developers are advised to individually consider the best evaluation data for their research.

\subsubsection{Class template}
\label{s:template}

To further simplify using the framework for developing new HRTF representations or adapting existing models, an \textsc{ooDir} class template is provided in file \texttt{ooDir\_template.m}. The name convention of an \textsc{ooDir} directivity model class is prefix \textit{ooDir\_} followed by an acronym of the method, with the exception of \texttt{RawIRs}, which has no prefix as the very basic representation. The template provides proper structure of the class as well as hints on how to fill it. With the introduction of the template, the entry level for using the framework was brought down significantly.

\subsubsection{Coordinate system converter}
\label{s:iap}

The entire toolbox employs exclusively standard spherical coordinate system, as depicted in Fig. \ref{f:coordsystem}. However, for some research, interaural-polar coordinate system introduced by Morimoto and Aokata in 1984 is deemed more appropriate \cite{Morimoto1984}. Thus, converters \texttt{sph2iap} and \texttt{iap2sph} were developed to freely move between vertical-polar (standard spherical) coordinate system and interaural-polar one. The converters are based on modernized definition of the interaural-polar system (e.g. \cite{Algazi2001}\cite{Romigh2015}), which is essentially the same as the spherical coordinate system employed in \textsc{ooDir} but rotated 90$\degree$ so that the poles are on the left and right side instead of at the zenith and at the bottom, respectively. The conversion is implemented in three steps:

\begin{enumerate}
\item{Convert to Cartesian coordinates using MATLAB function \texttt{sph2cart} with constant radius.}
\item{Rotate the Cartesian coordinates 90$\degree$ about x-axis:
	\begin{itemize}
	\item{counterclockwise for \texttt{sph2iap}, i.e. vertical-polar to interaural-polar coordinates}
	\item{clockwise for \texttt{iap2sph}, i.e. interaural-polar to vertical-polar coordinates}
	\end{itemize}}
\item{Convert rotated Cartesian coordinates back to spherical coordinate system using \texttt{cart2sph}.}
\end{enumerate}

\subsubsection{Difference class}
\label{s:diff}

During the objective evaluation process, the discrepancy between the model and the reference data is substantialized. Usually, it is calculated by averaging differences between values at individual datapoints. To facilitate this process, I developed a dedicated class \texttt{ooDir\_diff} for such comparisons, which is meant to store these individual differences and enable accessing them through the same interface as directivity representation classes.

As any \textsc{ooDir} class, \texttt{ooDir\_diff} is required to include properties \texttt{info} and \texttt{coords}. Besides them, there is a data matrix \texttt{diffM} storing the individual differences; the matrix has a typical structure of directions, frequencies and distances in consecutive dimensions. Furthermore, a reference object is included in the properties, as, apart from the differences themselves, the reference data is also needed for some error measures. Finally, there is a property \texttt{datatype}, which is a string specifying on what type of data the differences were calculated, e.g. complex or linear magnitude. The property itself is not public to avoid corruption, but its value is automatically appended to \texttt{info} in the constructor, so that anyone can read it.

The constructor has a very simple structure; as arguments, it takes \texttt{info}, two \textsc{ooDir} objects (a reference and an evaluand), \texttt{Coordinates} object and \texttt{datatype} string. Then, in the body of the constructor, the data of requested datatype is read from both objects at requested coordinates and subtracted. If there are any discrepancies between what is requested and what is available, the constructor issues appropriate errors or warnings.

By getting access to the base methods of \texttt{Directivity}, individual errors can be easily browsed and plotted. However, objective evaluation usually is focused on averaged errors. As described in section \ref{s:motivation}, there are many different ways of averaging differences, but two of them are especially common. One is already mentioned in section \ref{s:motivation} RMS of differences in dBs, which is usually called spectral distortion (SD) and can be defined as following:

\begin{equation}
\text{SD} = \sqrt{ \frac{1}{N} \sum\limits_{n=1}^{N} 20\log_{10}\left( \frac{|H[n]|}{| H_{\text{ref}}[n] |} \right)^{2} } \, .
\label{e:SD}
\end{equation}

\noindent where $H[n]$ and $H_{\text{ref}}[n]$ are $n$th elements of $N$-long vectors of evaluand and reference directivity values, respectively. The most popular alternative to SD is an error measure defined on the differences between linear (either absolute or complex) values, here referred to as mean-square-error (MSE)\footnote{The mentioned error measure can be met under different names and different units (e.g. final value might be a simple ratio \cite{Zhang2015}, percentage \cite{Zhang2009} or given in dBs \cite{Zhang2010}), but the underlying error formula is the same and they can be easily converted between each other.}:

\begin{equation}
\text{MSE} = \frac{\sum_{n=1}^{N} |H[n] - H_{\text{ref}}[n]|^{2}}{\sum_{n=1}^{N}|H_{\text{ref}}[n]|^{2}} \, .
\label{e:MSE}
\end{equation}

Even though SD seems to be more popular in recent papers, both can be commonly met (sometimes even both simultaneously \cite{Hugeng2017}) and so they were both implemented within the framework as methods of \texttt{ooDir\_diff}. The methods take \texttt{Coordinates} object as an argument to specify the datapoints for averaging and the error measures are determined according to formulae given in Eqs (\ref{e:SD}) and (\ref{e:MSE}). The error measures can be obtained only for specific datatypes: SD requires differences in logarithmic scale while MSE can be computed on complex or linear magnitude data. While there are also other error measures incompatible with SD and MSE (e.g. \cite{Liu2019}\cite{Huang2009}), they are far less established and so their implementation, if needed, is left for the framework users.

The results of objective evaluation are often provided in the form of plots. Depending on the characteristics of a model, various data visualizations might be desired, but there are two types of plots which are universal and can be commonly met in research papers. First of them is plotting error averaged over the sphere against frequency (e.g. \cite{Hu2019}\cite{Liu2019}\cite{Engel2022}) and the second is plotting error averaged over frequency for datapoints lying on the horizontal plane (e.g. \cite{Zhang2009}\cite{Romigh2015}\cite{Duraiswaini2004}). Both plotting functions (called \texttt{plotFrequency} and \texttt{plotHorizontal}, respectively) require a string specifying the type of averaging (SD, MSE or user-defined), and optionally can also take frequency range as an argument, e.g. to limit the analysis to specific bands (e.g. \cite{Andreopoulou2013}).

\section{Exemplary model evaluation}
\label{s:example}

To showcase the possibilities provided by the \textsc{ooDir} framework, typical evaluation procedures for an exemplary model are presented. The examples are based on test data \texttt{KEMAR.mat} and on the class \texttt{ooDir\_1Dbf}, which uses simple basis functions for spectrum approximation and which can be found in the \textsc{ooDir} repository \cite{Szwajcowski2022}. However, neither the evaluation data nor the model are of particular importance in this context, as the examples are meant to present the convenience of performing a HRTF model evaluation within the framework rather than specific results or methodology. The detailed description of syntax can be found in the documentation at \url{https://oodir.sourceforge.io}.

Let the accuracy of approximating HRTF spectra with first 16 functions of Fourier series be investigated. To create an \textsc{ooDir} object with such approximation, the following code is needed:

\begin{lstlisting}[style=Matlab-editor, basicstyle=\small, columns=fullflexible, frame=single]
load('KEMAR.mat', 'kemar') % reference, RawIRs object
info = 'Fourier series, 16 coefficients';
F = ooDir_1Dbf(info, kemar, 'Fourier', 16); % create an object
\end{lstlisting}

\subsection{Informal evaluation}
\label{s:informal}

The first step after performing initial computations is usually informal browsing of the results to see if everything worked as intended. In the case of HRTF models, it can be achieved by plotting exemplary spectra of the new investigated representation against the raw data. In \textsc{ooDir} it requires only a few lines of code:

\begin{lstlisting}[style=Matlab-editor, basicstyle=\small, columns=fullflexible, frame=single]
phi = 0; theta = 0; % direction straight ahead
kemar.plotSpectrum([phi, theta]) % plot reference spectra
hold on
F.plotSpectrum([phi, theta]) % plot approximated spectra
legend('raw', 'approximated', 'location', 'best')
\end{lstlisting}

\begin{figure}[h]
\centering
\includegraphics[width=\linewidth]{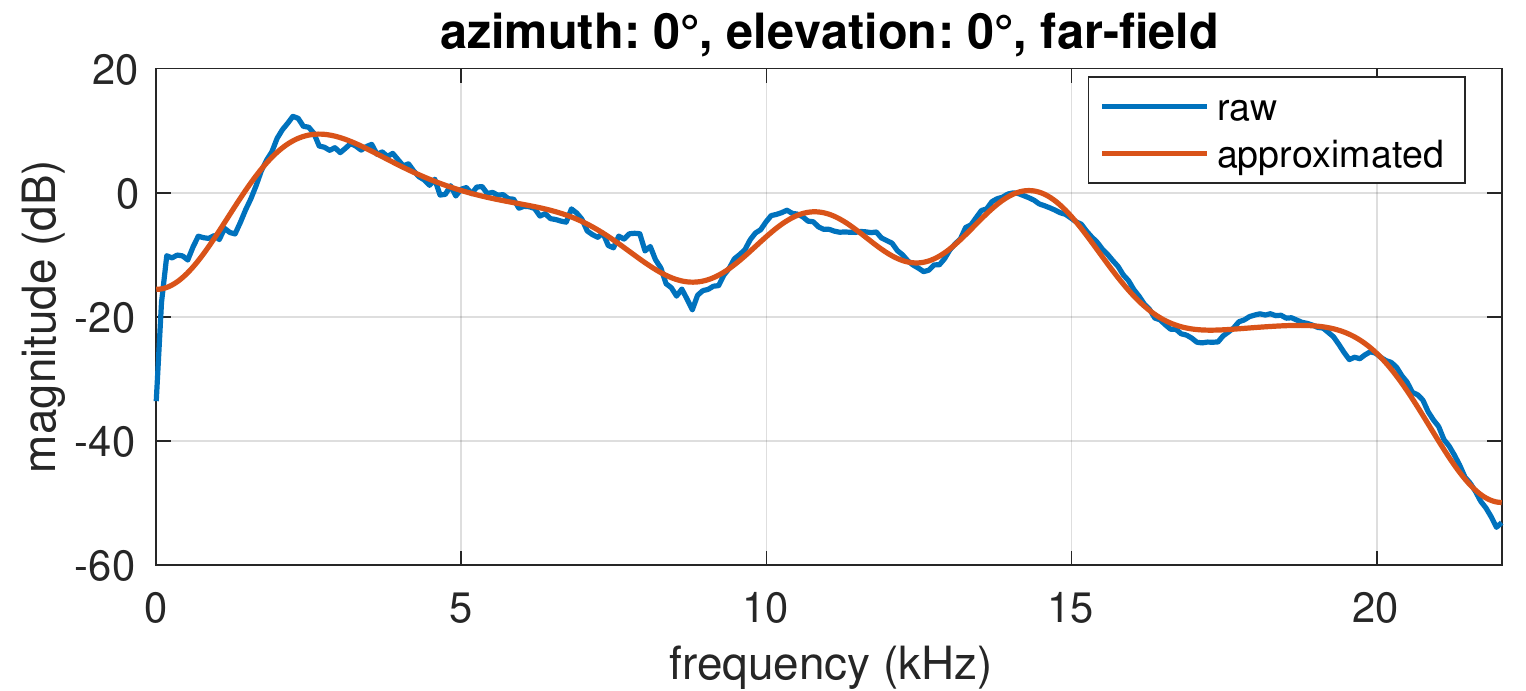}
\caption{Comparison of exemplary raw and approximated spectra}
\label{f:informal}
\end{figure}

\noindent The resulting plot is presented in Fig. \ref{f:informal}. Such result browsing can be useful not only for preliminary tests but also for capturing the character of approximations. Similar evaluation can be performed based on directivity balloons, although it is more suitable for sound source directivity rather than HRTFs.

\subsection{Averaged errors}
\label{s:average}

\begin{figure}[h]
\centering
\includegraphics[width=\linewidth]{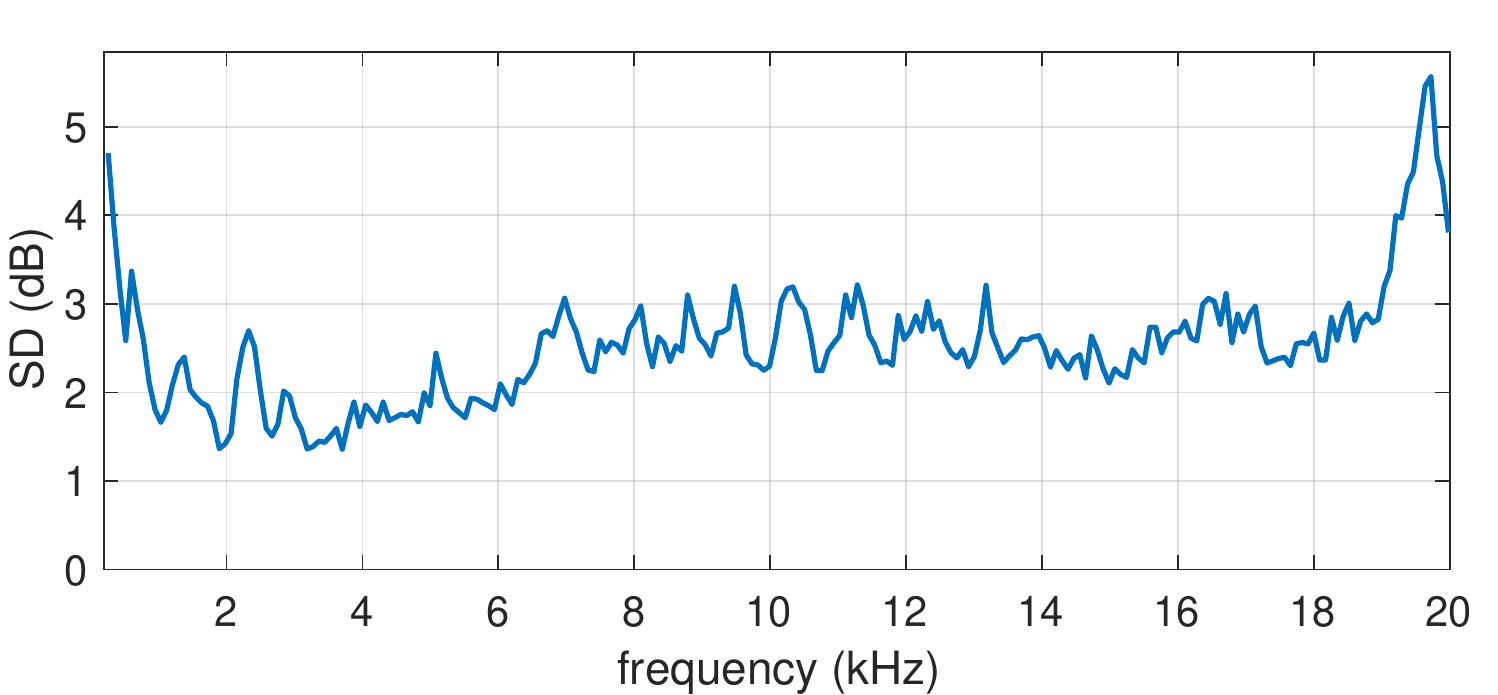}
\caption{Exemplary plot of averaged errors against frequency}
\label{f:average}
\end{figure}

Once it is established that the model is working fine, averaged errors can be calculated. Here, the \texttt{ooDir\_diff} class and its methods come useful for such analysis. For example, plotting the error averaged over the sphere against frequency can be obtained by typing following commands\footnote{As per documentation, leaving the \texttt{info} empty results in an automatically generated description}:

\begin{lstlisting}[style=Matlab-editor, basicstyle=\small, columns=fullflexible, frame=single]
Kcoords = kemar.getCoords; % coordinates for evaluation
D = ooDir_diff([], kemar, F, Kcoords, 'log'); % in dBs
D.plotFrequency('SD', [200, 20000]); % 200 Hz to 20 kHz
\end{lstlisting}

\noindent The results are presented in Fig. \ref{f:average}). A similar plot can be obtained for frequency-averaged error in the horizontal plane using the \texttt{plotHorizontal} method.

Another very popular type of figure is a plot of average errors depending on different values of model parameters. For example, for \texttt{ooDir\_1Dbf}, MSE can be determined for different number of Fourier series coefficients:

\begin{lstlisting}[style=Matlab-editor, basicstyle=\small, columns=fullflexible, frame=single]
N = 32; % maximum number of Fourier coefficients
MSE = zeros(N,1); % vector of MSE values
for n = 1:N
    F = ooDir_1Dbf(info, kemar, 'Fourier', n);
    D = ooDir_diff([], kemar, F, Kcoords, 'lin');
    % get MSE averaged over all frequencies and directions
    MSE(n) = D.computeMSE(Kcoords);
end
plot(MSE)
\end{lstlisting}

\subsection{Subjective evaluation}
\label{s:subjective}

Finally, the \textsc{ooDir} framework can also be useful in designing listening tests. Impulse responses at desired direction can be read from the test data by typing the following code:

\begin{lstlisting}[style=Matlab-editor, basicstyle=\small, columns=fullflexible, frame=single]
phi = 0; theta = 0; % direction straight ahead
IRcoords = Coordinates([], [phi,theta], [0,0,0])
% frequency field irrelevant for reading impulse responses
IR = kemar.getDataV(IRcoords,'IRs')
\end{lstlisting}

\noindent The obtained impulse response can be subsequently convoluted with a test signal. The presented code can work only with \textsc{ooDir} classes which support the 'IRs'  (impulse responses) datatype. The class \texttt{ooDir\_1Dbf} used in previous examples focuses only on magnitude modeling and thus it would return an error. To evaluate such model subjectively, phase should be included, e.g. based on interaural time differences \cite{Kulkarni2004}\cite{Romigh2015}. A new class would have to be developed (or the current one modified) to incorporate phase and thus enable extracting complex spectra or impulse responses.

Thanks to the common interface, all these pieces of code could be executed for just any \textsc{ooDir} class as long as it supports required datatypes. To obtain the results, one does not need to know the internal structure of the classes given that they trust its developer to properly implement the model and link it to the interface.

\section{Discussion}
\label{s:discussion}

The framework was successfully designed and implemented, fulfilling all imposed core features defined in section \ref{s:core}. It is \textbf{flexible}, allowing wide variety of models to be implemented and supporting even rare approaches such as time-based modeling or including distance for near-field directivity models. All representation methods are defined in the form of separate classes which are essentially single m-files for \textsc{Matlab}, thus exhibiting \textbf{modularity}. Furthermore, these classes inherit from the main class of the toolbox, \texttt{Directivity}, which provides a \textbf{common interface}. Finally, the entire interface has \textbf{fixed conventions and units}, operating in a single coordinate system (spherical, azimuth-elevation-distance) and using only SI units with the exception of angles utilizing degrees instead of radians as a more intuitive alternative.

Provided use-case examples showcase the simplicity of operating within the framework, as typical evaluation results can be obtained by just a couple of commands. One does not need to know the details of implementation as long as the classes are properly connected to the common interface, thus allowing for a quick and convenient comparison between different models. At the same time, conforming to the structure of an \textsc{ooDir} class is not problematic by itself and the toolbox includes an m-file template to make it even easier.

On the other hand, the framework also has some limitations. The primary one is that it is implemented as a \textsc{Matlab} toolbox, and so it can be only used by \textsc{Matlab} users. Due to its developmental nature, the framework cannot be precompiled and had to be written for a specific environment. \textsc{Matlab} was chosen as the most common programming environment in the audio engineering field, although it is not the only one. Since the toolbox does not include any commands exclusive to \textsc{Matlab}, it could also be rewritten in other languages; however, all the potential \textsc{ooDir} classes would also have to be converted (or written in more than one language by the authors themselves).

Another limitation is a somewhat strict data format, with output matrix structure being direction, frequency and distance dependencies in consecutive dimensions. Such format will work fine as long as the relative samplings are uniform. The problems could arise if e.g. for some directions frequency sampling was different than for the others. Although technically possible, there is no motivation behind varying the audio chain properties within the same set of measurements. The most probable scenario is different set of directions for different distances. Near-field directivity measurements (and thus models) are significantly less popular than the far-field ones and even then, near-field measurement usually do have uniform direction sampling for all distances (e.g. \cite{Brungart1999}\cite{Yu2018}). Furthermore, some workarounds are possible, such as creating multiple \textsc{ooDir} objects, each for data measured at different distance with non-uniform direction sampling.

Despite these potential drawbacks, the framework significantly enhances not only the development of new methods but also, even more importantly, the file communication. Thanks to the modular character of the toolbox, researchers from all over the world can conveniently share source code and test each others' models e.g. by evaluating them on different datasets or using different error measures. One could argue that it is better to pursue a unification of the form of results rather than creating a tool to support various evaluation procedures; however, at the point of writing this paper, the preferences are greatly varying among researchers and each approach has its own merits. A flexible framework for HRTF modeling such as \textsc{ooDir} is thus highly desirable and can contribute to finding a consensus in this matter, if it is ever to be worked out.

Furthermore, other types of spatially oriented data can also be modeled within \textsc{ooDir}, e.g. sound source directivity. Although such models are less popular than the HRTF ones, they are also investigated and could potentially be adapted or developed in the framework (e.g. \cite{Szwajcowski2021}\cite{Ahrens2021}).

Ultimately, the toolbox is expected to be complemented by a repository of \textsc{ooDir} classes and functions from directivity modeling research. Unifying the format of existing directivity models and those yet to be developed is a big step toward improving the comparability of alternative methods and thus could be an important contribution to determining the best ones for practical implementation.

\section{Conclusion}
\label{s:conclusion}

A framework for efficient developing and sharing HRTF models was successfully designed and implemented in the form of a \textsc{Matlab} toolbox under the name \textsc{ooDir}, standing for objective-oriented directivity. Each HRTF representation method is developed as a separate class inheriting from an abstract parent to maintain all the common features of a directivity function, while simultaneously leaving a lot of freedom as far as the exact definition of the given model is concerned. Being flexible and modular, as well as operating on fixed conventions and units and providing a common interface, it fills the niche of efficiently handling directivity data at the development stage. The toolbox requires only two classes, on top of which various directivity representations can be easily built. Both the core of the framework as well as supporting utility functions were explained and discussed. Additionally, typical evaluation procedures for an exemplary HRTF model were presented, highlighting the convenience of using the framework.

The greatest strength of \textsc{ooDir} is the simplicity of handling models with different definitions, even without the knowledge of the details of their implementation. Such feature allows the models developed within the framework to be easily shared and compared, thus pushing forward the entire field of HRTF representation models, by giving researchers a tool to validate each others' methods. Even for individual research, the framework is beneficial to use, as it provides a clean interface and convenient access to the data while being relatively simple to adopt. Finally, thanks to its universal implementation, it can be used not only for HRTFs but also for other types of directivity.

\section*{Acknowledgements}

This research was supported by the National Science Centre, project No. 2020/37/N/ST2/00122. The author would like to thank Anna Snakowska for her critical review of the manuscript before the submission.

\bibliographystyle{ieeetr}
\bibliography{bibJAES}

\begin{thebibliography}{10}

\bibitem{Kulkarni1995}
A.~Kulkarni and H.~S. Colburn, ``{Efficient finite‐impulse‐response filter
  models of the head‐related transfer function},'' {\em J. Acoust. Soc. Am.},
  vol.~97, pp.~3278--3278, May 1995.

\bibitem{Blommer1997}
M.~A. Blommer and G.~H. Wakefield, ``{Pole-zero approximations for head-related
  transfer functions using a logarithmic error criterion},'' {\em IEEE Trans.
  Speech Audio Process.}, vol.~5, pp.~278--287, May 1997.

\bibitem{Huopaniemi1999}
J.~Huopaniemi, N.~Zacharov, and M.~Karjalainen, ``{Objective and Subjective
  Evaluation of Head-Related Transfer Function Filter Design},'' {\em J. Audio
  Eng. Soc.}, vol.~47, pp.~218--239, April 1999.

\bibitem{Evans1998}
M.~J. Evans, J.~A.~S. Angus, and A.~I. Tew, ``{Analyzing head-related transfer
  function measurements using surface spherical harmonics},'' {\em J. Acoust.
  Soc. Am.}, vol.~104, pp.~2400--2411, October 1998.

\bibitem{Ziegler2017}
J.~D. Ziegler, M.~Rau, A.~Schilling, and A.~Koch, ``{Interpolation and Display
  of Microphone Directivity Measurements Using Higher Order Spherical
  Harmonics},'' in {\em 143rd AES Conv.}, Audio Engineering Society, October
  2017.

\bibitem{Szwajcowski2021}
A.~Szwajcowski, D.~Krause, and A.~Snakowska, ``{Error Analysis of Sound Source
  Directivity Interpolation Based on Spherical Harmonics},'' {\em Arch.
  Acoust.}, vol.~46, pp.~95--104, March 2021.

\bibitem{Zhang2009a}
W.~Zhang, T.~D. Abhayapala, R.~A. Kennedy, and R.~Duraiswami, ``{Modal
  expansion of HRTFs: Continuous representation in frequency-range-angle},'' in
  {\em IEEE Int. Conf. Acoust. Speech Signal Process.}, pp.~285--288, Institute
  of Electrical and Electronics Engineers Inc., May 2009.

\bibitem{Zhang2010}
W.~Zhang, T.~D. Abhayapala, R.~A. Kennedy, and R.~Duraiswami, ``{Insights into
  head-related transfer function: Spatial dimensionality and continuous
  representation},'' {\em J. Acoust. Soc. Am.}, vol.~127, pp.~2347--2357, apr
  2010.

\bibitem{Zhang2015}
M.~Zhang, R.~A. Kennedy, and T.~D. Abhayapala, ``{Empirical determination of
  frequency representation in spherical harmonics-based HRTF functional
  modeling},'' {\em IEEE/ACM Trans. Audio Speech Lang. Process.}, vol.~23,
  pp.~351--360, February 2015.

\bibitem{Bates2015}
A.~P. Bates, Z.~Khalid, and R.~A. Kennedy, ``{On the use of Slepian functions
  for the reconstruction of the head-related transfer function on the
  sphere},'' in {\em 9th Int. Conf. Signal Process. Commun. Syst. ICSPCS 2015},
  Institute of Electrical and Electronics Engineers Inc., December 2015.

\bibitem{Hu2019}
S.~Hu, J.~Trevino, C.~Salvador, S.~Sakamoto, and Y.~Suzuki, ``{Modeling
  head-related transfer functions with spherical wavelets},'' {\em Appl.
  Acoust.}, vol.~146, pp.~81--88, March 2019.

\bibitem{Liu2019}
H.~Liu, Y.~Fang, and Q.~Huang, ``{Efficient representation of head-related
  transfer functions with combination of spherical harmonics and spherical
  wavelets},'' {\em IEEE Access}, vol.~7, pp.~78214--78222, June 2019.

\bibitem{Kistler1992}
D.~J. Kistler and F.~L. Wightman, ``{A Model Of Head-Related Transfer Functions
  Based On Principal Components Analysis And Minimum-Phase Reconstruction},''
  {\em J. Acoust. Soc. Am.}, vol.~91, pp.~1637--1647, March 1992.

\bibitem{Huang2009}
Q.~Huang and K.~Liu, ``{A reduced order model of head-related impulse responses
  based on independent spatial feature extraction},'' in {\em IEEE Int. Conf.
  Acoust. Speech Signal Process.}, pp.~281--284, May 2009.

\bibitem{Shekarchi2013}
S.~Shekarchi, J.~Hallam, and J.~Christensen-Dalsgaard, ``{Compression of
  head-related transfer function using autoregressive-moving-average models and
  Legendre polynomials},'' {\em J. Acoust. Soc. Am.}, vol.~134, pp.~3686--3696,
  November 2013.

\bibitem{Arevalo2020}
C.~Arevalo and J.~Villegas, ``{Compressing Head-Related Transfer Function
  databases by Eigen decomposition},'' in {\em IEEE 22nd Int. Work. Multimed.
  Signal Process.}, Institute of Electrical and Electronics Engineers Inc.,
  September 2020.

\bibitem{Brinkmann2018}
F.~Brinkmann and S.~Weinzierl, ``{Comparison of head-related transfer functions
  pre-processing techniques for spherical harmonics decomposition},'' in {\em
  AES Int. Conf. Audio Virtual Augment. Real.}, August 2018.

\bibitem{Engel2022}
I.~Engel, D.~F.~M. Goodman, and L.~Picinali, ``Assessing hrtf preprocessing
  methods for ambisonics rendering through perceptual models,'' {\em Acta
  Acustica}, vol.~6, January 2022.

\bibitem{Majdak2013}
P.~Majdak, Y.~Iwaya, T.~Carpentier, R.~Nicol, M.~Parmentier, A.~Roginska,
  Y.~Suzuki, K.~Watanabe, H.~Wierstorf, H.~Ziegelwanger, and M.~Noisternig,
  ``{Spatially Oriented Format for Acoustics: A Data Exchange Format
  Representing Head-Related Transfer Functions},'' in {\em 134th AES Conv.},
  Audio Engineering Society, May 2013.

\bibitem{SOFA2020}
``{AES69-2020: AES standard for file exchange - Spatial acoustic data file
  format},'' tech. rep., Audio Engineering Society, June 2020.

\bibitem{Szwajcowski2021a}
A.~Szwajcowski, ``{Objective-oriented method for uniformation of various
  directivity representations},'' in {\em 151st AES Conv.}, (Las Vegas), Audio
  Engineering Society, October 2021.

\bibitem{Gardner1995}
W.~G. Gardner and K.~D. Martin, ``{HRTF measurements of a KEMAR},'' {\em J.
  Acoust. Soc. Am.}, vol.~97, pp.~3907--3908, June 1995.

\bibitem{Chen1995}
J.~Chen, B.~D. {Van Veen}, and K.~E. Hecox, ``{A spatial feature extraction and
  regularization model for the head-related transfer function},'' {\em J.
  Acoust. Soc. Am.}, vol.~97, pp.~439--452, January 1995.

\bibitem{Andreopoulou2013}
A.~Andreopoulou, A.~Roginska, H.~Mohanraj, and N.~York, ``{Analysis of the
  Spectral Variations in Repeated Head-related Transfer Function
  Measurements},'' in {\em Int. Conf. Audit. Disp.}, pp.~213--218, Georgia
  Institute of Technology, July 2013.

\bibitem{Zhang2009}
W.~Zhang, R.~A. Kennedy, and T.~D. Abhayapala, ``{Efficient continuous HRTF
  model using data independent basis functions: Experimentally guided
  approach},'' {\em IEEE Trans. Audio, Speech Lang. Process.}, vol.~17,
  pp.~819--829, May 2009.

\bibitem{Kulkarni2004}
A.~Kulkarni and H.~S. Colburn, ``{Infinite-impulse-response models of the
  head-related transfer function},'' {\em J. Acoust. Soc. Am.}, vol.~115,
  pp.~1714--1728, April 2004.

\bibitem{Iida2007}
K.~Iida, M.~Itoh, A.~Itagaki, and M.~Morimoto, ``{Median plane localization
  using a parametric model of the head-related transfer function based on
  spectral cues},'' {\em Appl. Acoust.}, vol.~68, pp.~835--850, August 2007.

\bibitem{Li2021}
J.~Li, B.~Wu, D.~Yao, and Y.~Yan, ``A mixed-order modeling approach for
  head-related transfer function in the spherical harmonic domain,'' {\em
  Applied Acoustics}, vol.~176, January 2021.

\bibitem{Hrauda2013}
W.~Hrauda, ``Essentials on hrtf measurement and storage format
  standardization,'' bachelor thesis, Graz University of Technology, 2013.

\bibitem{Diemo2000}
D.~Schwarz and M.~Wright, ``Extensions and applications of the sdif sound
  description interchange format,'' in {\em International Computer Music
  Conference ICMC2000}, August 2000.

\bibitem{Wefers2010}
F.~Wefers, ``Opendaff a free, open-source software package for directional
  audio data,'' in {\em DAGA}, March 2010.

\bibitem{Dietrich2012}
P.~Dietrich, M.~Guski, M.~Pollow, M.~M{\"u}ller-Trapet, B.~Masiero,
  R.~Scharrer, and M.~Vorl{\"a}nder, ``Ita-toolbox-an open source matlab
  toolbox for acousticians,'' in {\em DAGA}, March 2012.

\bibitem{Morimoto1984}
M.~Morimoto and H.~Aokata, ``Localization cues of sound sources in the upper
  hemisphere,'' {\em J. Acoust. Soc. Jap.}, vol.~5, no.~3, pp.~165--173, 1984.

\bibitem{Algazi2001}
V.~R. Algazi, R.~O. Duda, D.~M. Thompson, and C.~Avenda{\~{n}}o, ``{The CIPIC
  HRTF database},'' in {\em IEEE ASSP Work. Appl. Signal Process. to Audio
  Acoust.}, pp.~99--102, October 2001.

\bibitem{Romigh2015}
G.~D. Romigh, D.~S. Brungart, R.~M. Stern, and B.~D. Simpson, ``{Efficient Real
  Spherical Harmonic Representation of Head-Related Transfer Functions},'' {\em
  IEEE J. Sel. Top. Signal Process.}, vol.~9, pp.~921--930, August 2015.

\bibitem{Hugeng2017}
H.~Hugeng, J.~Anggara, and D.~Gunawan, ``Enhanced three-dimensional hrirs
  interpolation for virtual auditory space,'' in {\em IEEE Int. Conf. Signals
  Syst. ICSigSys 2017}, pp.~35--39, May 2017.

\bibitem{Duraiswaini2004}
R.~Duraiswaini, D.~Zotkin, and N.~Gumerov, ``{Interpolation and range
  extrapolation of HRTFs},'' in {\em IEEE Int. Conf. Acoust. Speech Signal
  Process.}, pp.~45--48, Institute of Electrical and Electronics Engineers
  (IEEE), September 2004.

\bibitem{Szwajcowski2022}
A.~Szwajcowski, ``Approximating head-related transfer functions in the domain
  of common basis functions,'' in {\em 28th Int. Congr. Sound Vib. ICSV28},
  July 2022.

\bibitem{Brungart1999}
D.~S. Brungart and W.~M. Rabinowitz, ``{Auditory localization of nearby
  sources. Head-related transfer functions},'' {\em J. Acoust. Soc. Am.},
  vol.~106, pp.~1465--1479, September 1999.

\bibitem{Yu2018}
G.~Yu, R.~Wu, Y.~Liu, and B.~Xie, ``{Near-field head-related transfer-function
  measurement and database of human subjects},'' {\em J. Acoust. Soc. Am.},
  vol.~143, pp.~EL194--EL198, March 2018.

\bibitem{Ahrens2021}
J.~Ahrens and S.~Bilbao, ``Computation of spherical harmonic representations of
  source directivity based on the finite-distance signature,'' {\em IEEE/ACM
  Trans. Audio Speech Lang. Process.}, vol.~29, pp.~83--92, 2021.

\end{thebibliography}

\end{document}